\documentclass[12pt]{iopart}
\pdfoutput=1
\usepackage{mathrsfs, bm, amssymb}
\usepackage{graphicx,color}

\usepackage{iopams}
\def\ra{\rangle}
\def\la{\langle}

\def\be{\beta}
\def\al{\alpha}

\def\sig{\sigma}
\def\bege{\begin{equation}}
\def\ende{\end{equation}}
\def\begen{\begin{eqnarray}}
\def\enden{\end{eqnarray}}

\begin{document}

\title[]{
On integrable matrix product operators with bond dimension $D=4$ 
}

\author{Hosho Katsura$^{1,2}$}

\address{${}^1$Department of Physics, Graduate School of Science, University of Tokyo, 7-3-1 Hongo, Bunkyo-ku, Tokyo 113-0033
}
\address{${}^2$Department of Physics, Gakushuin University, 1-5-1 Mejiro, Toshima-ku, Tokyo 171-8588, Japan}
\begin{abstract}
We construct and study a two-parameter family of matrix product operators of bond dimension $D=4$. The operators $M(x,y)$ act on $({\mathbb C}_2)^{\otimes N}$, i.e., the space of states of a spin-$1/2$ chain of length $N$. For the particular values of the parameters: $x=1/3$ and $y=1/\sqrt{3}$, the operator turns out to be proportional to the square root of the reduced density matrix of the valence-bond-solid state on a hexagonal ladder. 
We show that $M(x,y)$ has several interesting properties when $(x,y)$ lies on the unit circle centered at the origin: $x^2 + y^2=1$. In this case, we find that $M(x,y)$ commutes with the Hamiltonian and all the conserved charges of the isotropic spin-$1/2$ Heisenberg chain. Moreover, $M(x_1,y_1)$ and $M(x_2,y_2)$ are mutually commuting if $x^2_i + y^2_i=1$ for both $i=1$ and $2$. 
These remarkable properties of $M(x,y)$ are proved as a consequence of the Yang-Baxter equation.
\end{abstract}

% 02.30.Ik Integrable systems 
% 75.10.Kt	Quantum spin liquids, valence bond phases and related phenomena
% 75.10.Pq	Spin chain models
%\noindent{\it Keywords}: Matrix Product States, Algebraic Bethe Ansatz
%Article preparation, IOP journals
%\maketitle
%%%%%   Introduction   %%%%%

\section{Introduction}
\label{sec:intro}

Since the proposal by Li and Haldane~\cite{Li_Haldane}, the concept of entanglement spectrum has attracted much attention, 
particularly for studying topologically ordered systems. The entanglement spectrum is the eigenvalue spectrum 
of the entanglement Hamiltonian $H_{\rm E}$, the levels of which are in one-to-one correspondence with the eigenvalues of the reduced density matrix $\rho_A$ for a subsystem $A$. More precisely, $\rho_A$ is written in terms of $H_{\rm E}$ as $\rho_A = \exp(-H_{\rm E})$. To date, a number of studies on entanglement spectra have been carried out for a variety of systems. 
Examples include quantum Hall systems~\cite{Regnault, Zozulya, Lauchili, Thomale, Chandran, Qi_Ludwig}, topological insulators~\cite{Turner, Fidkowski, Prodan}, and quantum spin models in one~\cite{Calabrese, Pollmann1, Pollmann2, Poilblanc, Thomale2, Lauchli2} and two~\cite{Yao, Poilblanc1, Huang, Lou_PRB, Tanaka1} dimensions. 

In Ref. \cite{Tanaka_Tamura}, the author and his collaborators have studied a class of tensor-network states on ladders. The quantum lattice-gas model in which these states are the exact ground states was first introduced in Refs.~\cite{Lesanovsky1, Ates}. The model is referred to as the quantum hard-square model since the basis states of the Hilbert space are in one-to-one correspondence with allowed configurations of the classical hard-square model, i.e., any pair of adjacent sites cannot be occupied by more than two particles. 
The ground state can be expressed as a superposition of classical configurations, each of which has a weight depending on the parameter $z$. The meaning of $z$ in the classical model is the activity, the Boltzmann weight per particle. In Ref. \cite{Tanaka_Tamura}, the authors found that the entanglement spectrum of the ground state is critical for both square and triangular ladders when $z$ is chosen so that the corresponding classical model is critical. It was also revealed numerically that entanglement Hamiltonians for the square and triangular ladders are well described by $c=1/2$ and $c=4/5$ conformal field theories, respectively. 

In addition to critical properties, the authors found that the entanglement Hamiltonian as well as the reduced density matrix for a triangular ladder are integrable for arbitrary $z$ despite the fact that the physical Hamiltonian is non-integrable, i.e., we cannot obtain the eigenstates analytically except the ground state. Here, integrability means that reduced density matrices for different values of $z$ are mutually commuting, which implies the existence of an infinite number of conserved charges that commute with the entanglement Hamiltonian in infinite systems. This remarkable fact was proved by exploiting the close connection between the reduced density matrix for the triangular ladder and the transfer matrix of the classical hard-hexagon model, which is integrable and was solved by Baxter~\cite{Baxter_book}. For a detailed proof, see Appendix B in Ref.~\cite{Tanaka_Tamura}. 

In this paper, we introduce and study a two-parameter family of matrix product operators (MPOs) that exhibit similar integrable structures for special cases.  
MPOs are operator analogues of matrix product states~\cite{Verstraete1, Vidal1}, which have recently attracted much attention from both quantum information and condensed matter points of view. In fact, our construction of the MPOs is motivated by the recent study of entanglement spectra of valence-bond-solid (VBS) states in two dimensions~\cite{Poilblanc1, Lou_PRB, KKKKT, Santos}. Note that the VBS states are the exact ground states of the Affleck-Kennedy-Lieb-Tasaki (AKLT) model~\cite{AKLT1, AKLT2, KLT}, and can also be described as tensor network states. We will show that the MPO constructed, say $M(x,y)$, is proportional to the square root of the reduced density matrix of the VBS state on a hexagonal ladder~\cite{KKKKT} for the particular values of the parameters: $x=1/3$ and $y=1/\sqrt{3}$. Unfortunately, the MPOs with this particular set of parameters do not exhibit any integrability property. However, 
we find a surprising property of the MPOs for other special cases. If $(x, y)$ are on the unit circle centered at the origin, i.e., $x^2+y^2=1$, 
the operators $M(x,y)$ commute with the transfer matrix of the  isotropic spin-$1/2$ Heisenberg chain, which is usually called as the XXX chain. 

The XXX chain is a prototypical example of quantum integrable models and its integrability is well explained by the existence of a family of commuting transfer matrices. They can be constructed in the framework of the algebraic Bethe ansatz (ABA)~\cite{Faddeev, Korepin_book, Nepomechie}. From a modern point of view,  these transfer matrices can also be thought of as MPOs. The relation between the ABA and matrix product states was discussed in detail in Refs. \cite{Katsura_Maruyama1, Katsura_Maruyama2, Murg1}. It is interesting to note that the bond dimension of the $M(x,y)$ is $D=4$, while that of the transfer matrix of the XXX chain is $D=2$. 
Recently, MPOs with finite ($D=4$) or infinite bond dimension ($D=\infty$) that exhibit some integrable structure have proved to be useful in describing non-equilibrium steady states of anisotropic Heisenberg chains driven by local noise at their boundaries~\cite{Znidaric1, Prosen1, Prosen2, Karevski, Prosen3}. 

In this paper, we prove that $M(x,y)$ with $(x,y)$ being on the unit circle centered at the origin is proportional to the product of two transfer matrices of the XXX chain with different spectral parameters. Then, combining this fact with the Yang-Baxter equation, we show that in this case $M(x,y)$ commute with the transfer matrix of the XXX chain, which readily implies that $M(x,y)$ commute with the Heisenberg Hamiltonian $H$ and all other conserved charges derived from the logarithmic derivatives of the transfer matrix. 
In addition, we also show that $M(x_1,y_1)$ and $M(x_2,y_2)$ are mutually commuting if $x^2_i+y^2_i=1$ for both $i=1$ and $2$. 
For small system sizes ($N \le 5$), these properties hold even for arbitrary $(x,y)$, which can be verified by explicitly writing out $M(x,y)$ as polynomials in $H$ and $({\bm \sigma}_{\rm tot})^2$, with ${\bm \sigma}_{\rm tot}$ being the total spin operator. 

The rest of the paper is organized as follows. In section 2, we first give a general definition of MPOs of bond dimension $D$, acting on the space of states of a spin-$1/2$ chain. Then, in section 2.1, we briefly review the construction of the transfer matrix of the XXX chain through the ABA. In section 2.2, we introduce $M(x,y)$, the MPOs of $D=4$ with the two parameters $x$ and $y$. We show that $M(x,y)$ is related to the VBS state on a hexagonal ladder if $x=1/3$ and $y=1/\sqrt{3}$. We also show that $M(x,y)$ can be thought of as a two-parameter deformation of the Gram matrix associated with this VBS state. In section 3, we discuss integrability properties of $M(x,y)$ for special cases where $(x,y)$ lie on the unit circle centered at the origin. We show that these integrability properties can be proved as a consequence of the Yang-Baxter equation. 
Concluding remarks are given in the last section. 
In Appendix A, the explicit expressions for $M(x,y)$ up to $N=5$ are presented. 

%%%%%%%%%%%%%%%%%%%%%%%%%%%%%%%%%%%%%%%%%%%%%%%%%%%%%%%%%

\section{Matrix product operators}
\label{sec:MPO}
\begin{figure}[htb]
\begin{center}
\vspace{.5cm}
\hspace{-.0cm}\includegraphics[width=0.95\columnwidth]{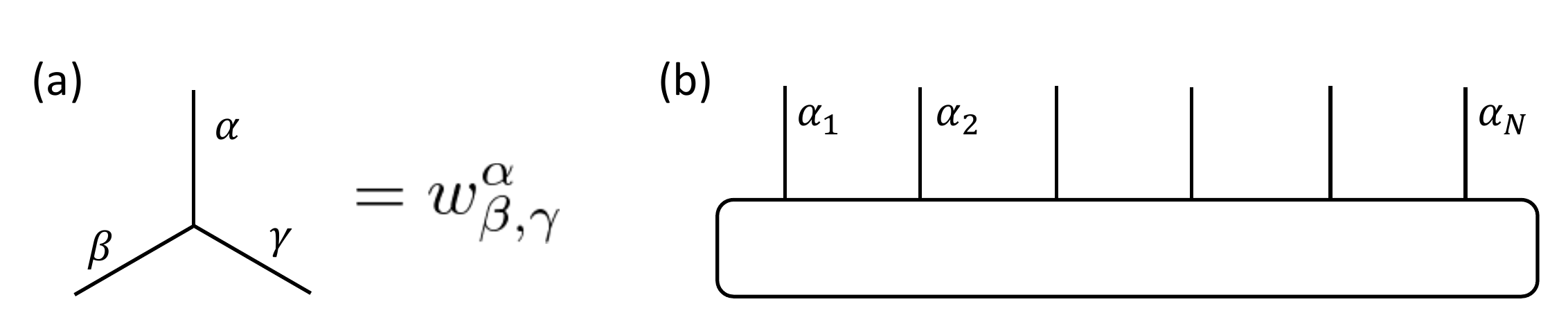}
\vspace{.0cm}
\caption{(a) Boltzmann weights for a trivalent vertex. (b) Comb lattice and a graphical expression for ${\rm Tr}[ {\sf W}^{\al_1} {\sf W}^{\al_2} \cdots {\sf W}^{\al_N} ]$. The degrees of freedom live on the edges of the lattice.}
\label{fig: weight1}
\end{center}
\end{figure}

We start with a general definition of matrix product operators acting on ${\mathscr H}_N=({\mathbb C}_2)^{\otimes N}$, i.e., the space of states of a spin-$1/2$ chain with $N$ sites. In the following, we will use the standard notation to denote the identity and Pauli matrices:
\begin{eqnarray}
\!\!\!\!\!\!\!\!\!\!\!\!\!\!\!
\sig^0 = \left(\begin{array}{cc}1 & 0 \\ 0 & 1 \end{array}\right),~~
\sig^1 = \left(\begin{array}{cc}0 & 1 \\ 1 & 0 \end{array}\right),~~
\sig^2 = \left(\begin{array}{cc}0 & -i \\ i & 0 \end{array}\right),~~
\sig^3 = \left(\begin{array}{cc}1 & 0 \\ 0 & -1 \end{array}\right).  
\end{eqnarray}
A translation invariant MPO acting on ${\mathscr H}_N$ is written in the form
\begin{equation}
W = \sum^3_{\al_1, ..., \al_N=0} 
{\rm Tr} [ {\sf W}^{\al_1} {\sf W}^{\al_2} \cdots {\sf W}^{\al_N} ]\,
\sig^{\al_1} \otimes \sig^{\al_2} \otimes \cdots \otimes \sig^{\al_N},
\label{eq:MPO1}
\end{equation}
where ${\sf W}^\al$ ($\al=0,1,2,3$) are matrices of bond dimension $D$. The matrix elements $({\sf W}^\al)_{\beta, \gamma}=w^\al_{\beta,\gamma}$ ($\beta, \gamma=0,1, ..., D-1$) can be thought of as Boltzmann weights for the vertex model on a comb  (see Fig. \ref{fig: weight1}). Any translation invariant operator acting on ${\mathscr H}_N$ can, in principle, be written in the form of equation (\ref{eq:MPO1}). In the following, we first show that the transfer matrix of the XXX chain can be written in the form of equation (\ref{eq:MPO1}). 
Then we introduce a two-parameter family of MPOs, which are not identical to the transfer matrix. The MPOs introduced are related to a deformation of the reduced density matrix of the VBS state on a hexagonal ladder. 

\subsection{Transfer matrix of the XXX chain}
From the algebraic Bethe ansatz~\cite{Faddeev, Korepin_book, Nepomechie}, one can see that the transfer matrix of the XXX chain with periodic boundary conditions can be cast into the form equation (\ref{eq:MPO1}): 
\begin{equation}
T (\lambda) = \sum^3_{\al_1, ..., \al_N=0} 
{\rm Tr} [ {\sf L}^{\al_1}(\lambda) {\sf L}^{\al_2}(\lambda) \cdots {\sf L}^{\al_N}(\lambda) ]\,
\sig^{\al_1} \otimes \sig^{\al_2} \otimes \cdots \otimes \sig^{\al_N},
\label{eq:trmat}
\end{equation}
where ${\sf L}^0 (\lambda) = \lambda \sig^0$ and ${\sf L}^a (\lambda) = i \sig^a/2$ ($a=1,2,3$) are matrices of bond dimension $D=2$. Here $\lambda$ is an arbitrary parameter, called a spectral parameter. The operator ${\cal L}(\lambda) := \sum^3_{\al=0} {\sf L}^\al (\lambda) \otimes \sig^\al$ is called the $L$-operator in the literature of quantum integrable systems. 
The central object in the ABA is the quantum $R$-matrix, which is a solution of the Yang-Baxter equation. For the XXX chain, it is given by
\begin{equation}
{\sf R} (\lambda) = 
\left(\begin{array}{cccc} 
\lambda+i & 0 & 0 & 0 \\
0 & \lambda & i & 0 \\
0 & i & \lambda & 0 \\
0 & 0 & 0 & \lambda+i
\end{array}\right).
\end{equation}
By direct calculation, one can verify the following identity ($RLL=LLR$ relation):
\begin{equation}
({\sf R} (\lambda-\mu) \otimes \sig^0)\, {\cal L}_1 (\lambda) {\cal L}_2 (\mu) 
= {\cal L}_2 (\mu) {\cal L}_1 (\lambda)\, ({\sf R} (\lambda-\mu) \otimes \sig^0),
\label{eq:RLL}
\end{equation}
where 
\begin{equation}
{\cal L}_1 (\lambda) = \sum^3_{\alpha=0} {\sf L}^\alpha (\lambda) \otimes \sig^0 \otimes \sig^\alpha,~~~
{\cal L}_2 (\mu) = \sum^3_{\alpha=0} \sig^0 \otimes {\sf L}^\alpha (\mu) \otimes \sig^\alpha
\end{equation}
It follows from Eq. (\ref{eq:RLL}) that $T(\lambda)$ is a one-parameter family of commuting matrices, i.e., 
\begin{equation}
[T(\lambda), T(\mu)]=0,
\label{eq:TT}
\end{equation} 
(for a proof, see Refs.~\cite{Faddeev, Korepin_book, Nepomechie}). 
The commutativity of transfer matrices implies that the logarithmic derivatives of $T(\lambda)$ are mutually commuting. An example is the first derivative with respect to $\lambda$:
\begin{eqnarray}
2i \frac{d}{d\lambda} \ln T(\lambda) \bigg|_{\lambda=i/2} = H + {\rm const.}
\end{eqnarray}
Here $H$ is the Hamiltonian for the XXX chain:
\begin{equation}
H = \sum^N_{j=1} (\sigma^1_{j} \sigma^1_{j+1} + \sigma^2_{j} \sigma^2_{j+1} + \sigma^3_{j} \sigma^3_{j+1} ),
\end{equation}
where $\sig^a_j = (\bigotimes^{j-1}_{k=1} \sigma^0) \otimes \sig^a (\bigotimes^{N-j}_{k=1} \sigma^0 )$ ($a=1,2,3$) and the periodic boundary conditions are imposed, i.e., $\sigma^a_{N+1} = \sigma^a_1$. 
The other conserved charges (higher Hamiltonians) can also be obtained via
\begin{eqnarray}
Q_n := 2i \frac{d^{n-1}}{d\lambda^{n-1}} \ln T(\lambda) \bigg|_{\lambda=i/2}.
\label{eq:higher_charge}
\end{eqnarray}
The conserved charges so obtained are local operators, i.e., the range of interactions does not increase with the size of the chain. However, their ``explicit" expressions are quite complicated~\cite{Grabowski1, Grabowski2, Grabowski3}. Note that for infinite chains, there is a shortcut. One can obtain $Q_n$ recursively by taking the commutator with the boost operator~\cite{Tetelman, Thacker, Sklyanin}. 
%%%%%%%%%%%%%%%%%%%%%%%%%%%%%%%%%%%%%%

\subsection{Matrix product operators $M(x,y)$}

\begin{figure}[htb]
\begin{center}
\vspace{.5cm}
\hspace{-.0cm}\includegraphics[width=0.95\columnwidth]{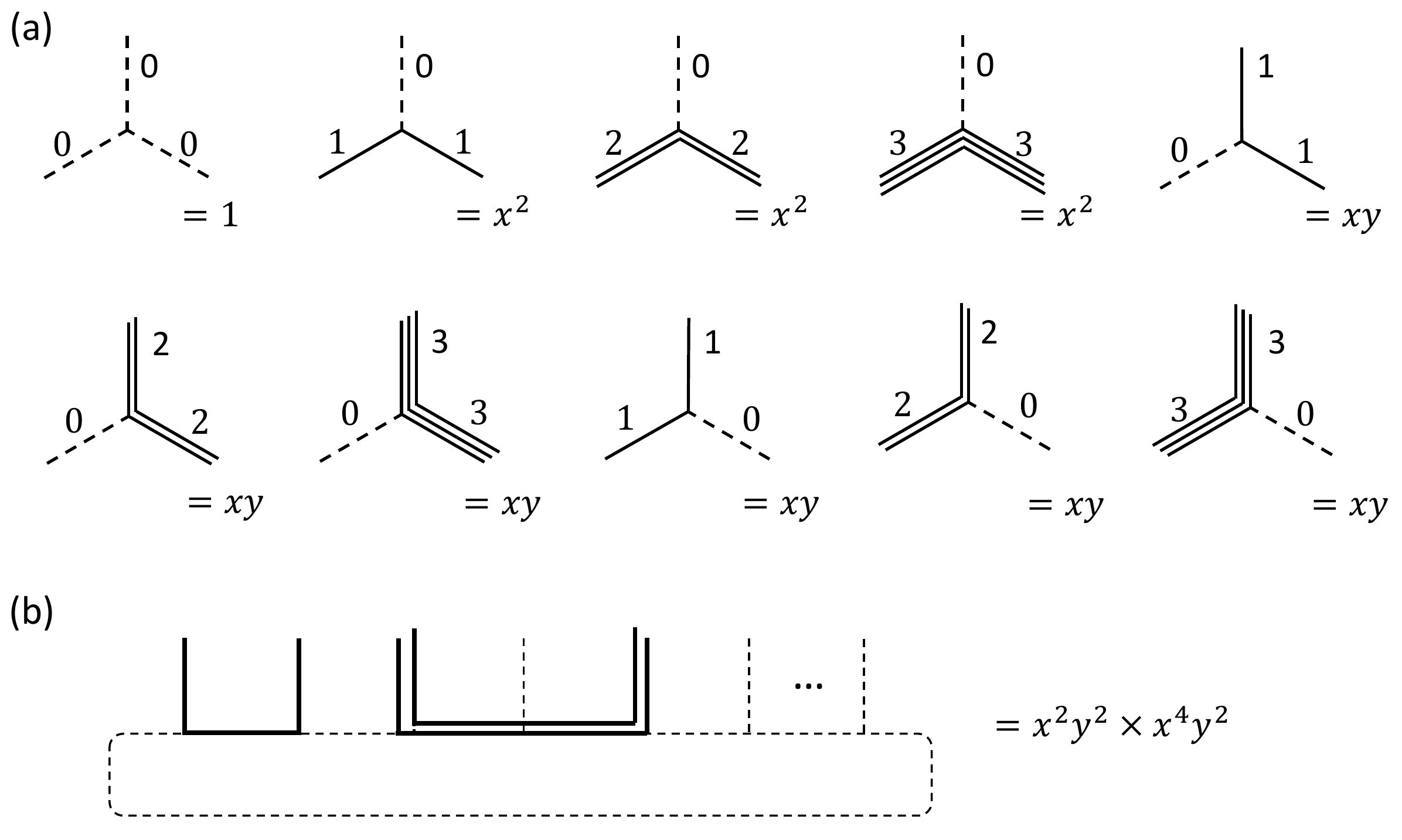}
\vspace{.0cm}
\caption{(a) Vertex configurations with nonvanishing Boltzmann weights. (b) The diagram corresponding to the coefficient of $\sigma^1_1 \sigma^1_2 \sigma^2_3 \sigma^2_5$ in $M(x,y)$. }
\label{fig: weights}
\end{center}
\end{figure}

Now, let us introduce the following matrix product operators with bond dimension $D=4$:
\begin{equation}
\!\!\!\!\!\!\!\!\!\!
M (x,y) = \sum^3_{\al_1, ..., \al_N=0} 
{\rm Tr} [ {\sf M}^{\al_1}(x,y) {\sf M}^{\al_2}(x,y) \cdots {\sf M}^{\al_N}(x,y) ]\,
\sig^{\al_1} \otimes \sig^{\al_2} \otimes \cdots \otimes \sig^{\al_N},
\label{eq:cons}
\end{equation}
where the matrices depending on the parameters $x$ and $y$ are given by
\begin{eqnarray}
\!\!\!\!\!
{\sf M}^0 (x,y) = 
\left(\begin{array}{cccc} 
1 & 0 & 0 & 0 \\
0 & x^2 & 0 & 0 \\
0 & 0 & x^2 & 0 \\
0 & 0 & 0 & x^2
\end{array}\right),~~
{\sf M}^1 (x,y) = 
\left(\begin{array}{cccc} 
0 & xy & 0 & 0 \\
xy & 0 & 0 & 0 \\
0 & 0 & 0 & 0 \\
0 & 0 & 0 & 0
\end{array}\right),
\nonumber \\
\!\!\!\!\!
{\sf M}^2 (x,y) = 
\left(\begin{array}{cccc} 
0 & 0 & xy & 0 \\
0 & 0 & 0 & 0 \\
xy & 0 & 0 & 0 \\
0 & 0 & 0 & 0
\end{array}\right),~~
{\sf M}^3 (x,y) = 
\left(\begin{array}{cccc} 
0 & 0 & 0 & xy \\
0 & 0 & 0 & 0 \\
0 & 0 & 0 & 0 \\
xy & 0 & 0 & 0
\end{array}\right).
\label{eq: Mmat1}
\end{eqnarray}
Using $\{ |0\ra, |1\ra, |2\ra, |3\ra \}$, an orthonormal basis of the auxiliary space, 
the above matrices can be expressed more compactly as ${\sf M}^0 (x,y) =| 0 \ra \la 0| + x^2 \sum^3_{a=1} |a\ra \la a|$,
and ${\sf M}^a (x,y) = xy | 0 \ra \la a | + xy| a \ra \la 0|$ ($a=1,2,3$). The matrix elements $({\sf M^\al})_{\beta,\gamma}$ can again be interpreted as Boltzmann weights for the vertex model on a comb. The configurations with nonvanishing Boltzmann weights are shown in Fig. \ref{fig: weights} (a). From the graphical rules, the coefficient of each term in $M(x,y)$ can be easily read off. 
An example is shown in Fig. \ref{fig: weights} (b). 

\begin{figure}[htb]
\begin{center}
\vspace{.5cm}
\hspace{-.0cm}\includegraphics[width=0.85\columnwidth]{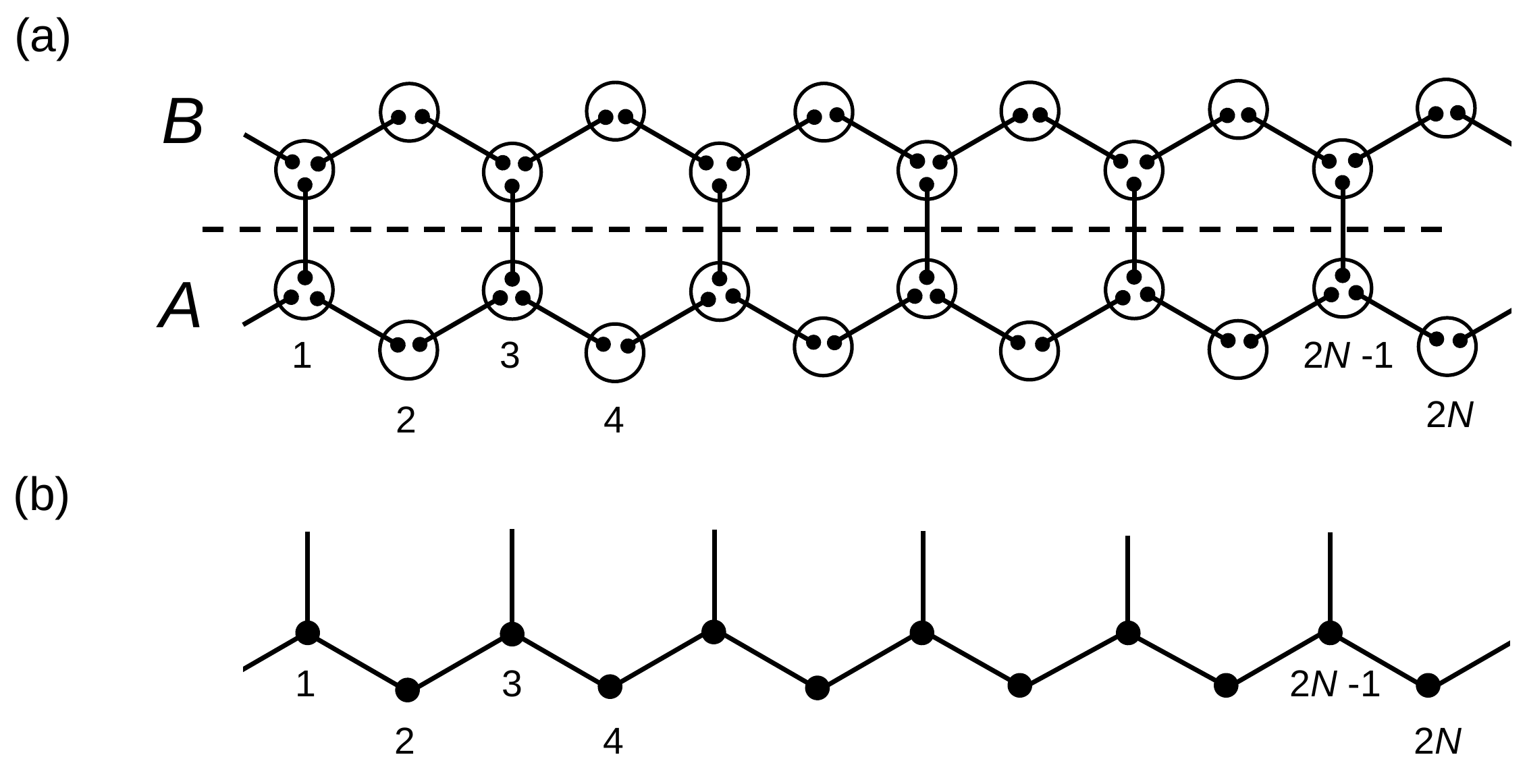}
\vspace{.0cm}
\caption{(a) Valence-bond-solid state on a hexagonal ladder. The boundary between  subsystems $A$ and $B$ is indicated by the broken line. (b) Comb lattice with vertices $k=1,2,...,2N-1, 2N$.}
\label{fig: VBS}
\end{center}
\end{figure}

The specific form of $M(x,y)$ is motivated by the recent study of entanglement spectra of VBS states on hexagonal lattices~\cite{Poilblanc1, Lou_PRB, Santos}. For a hexagonal ladder shown in Fig. \ref{fig: VBS} (a), the spectrum of the reduced density matrix of subsystem $A$ is identical to that of the following matrix:
\begin{equation}
{\hat \rho}_A = \frac{(M_{\rm VBS})^2}{{\rm Tr} [(M_{\rm VBS})^2]},
\end{equation}
where the Gram matrix $M_{\rm VBS}$ is defined by
\begin{equation}
M_{\rm VBS} = \int \left( \prod^{2N}_{k=1} \frac{d {\bm \Omega}_k}{4 \pi} \right)
\prod^N_{j=1} (1+{\bm \sigma}_j \cdot {\bm \Omega}_{2j-1}) 
\prod^{2N}_{k=1} (1+{\bm \Omega}_k \cdot {\bm \Omega}_{k+1}). 
\label{eq:MVBS}
\end{equation}
Here ${\bm \Omega}_k$ is the unit vector defined by 
${\bm \Omega}_k := ( \sin\theta_k \cos \phi_k, \sin\theta_k \sin \phi_k,\cos\theta_k)$ 
and ${\bm \sigma}_j := (\sigma^1_j, \sigma^2_j, \sigma^3_j)$ ($j=1, 2, ..., N$) are the Pauli matrix vectors. The periodic boundary conditions imply ${\bm \sigma}_{N+1} = {\bm \sigma}_1$ and ${\bm \Omega}_{2N+1} = {\bm \Omega}_1$. Note that the integral expression Eq. (\ref{eq:MVBS}) was derived using Schwinger bosons and the spin coherent state representation \cite{Arovas_PRL} (for derivation, see Refs. \cite{Lou_PRB, KKKKT}). 
The MPO expression of $M_{\rm VBS}$ can be derived by multiplying out the factors in the product in equation (\ref{eq:MVBS}) and carrying out the integrals over ${\bm \Omega}_k$. 
In each term in the sum, ${\bm \Omega}_k$ appears zero, one, two, or three times for each $k$. By symmetry, or by direct calculation, one finds the following rules:
\begin{eqnarray}
\int \frac{d {\bm \Omega}_k}{4\pi}\, \Omega^a_k = 0, ~~~
\int \frac{d {\bm \Omega}_k}{4\pi}\, \Omega^a_k \, \Omega^b_k = \frac{1}{3}\delta^{ab}, ~~~
\int \frac{d {\bm \Omega}_k}{4\pi}\, \Omega^a_k \, \Omega^b_k \, \Omega^c_k = 0
\label{eq:rules}
\end{eqnarray}
($a,b,c=1,2,3$), from which it immediately follows that 
\begin{equation}
\int \frac{d {\bm \Omega}_k}{4\pi} ({\bm v}_j \cdot {\bm \Omega}_k) ({\bm v}_\ell \cdot {\bm \Omega}_k) = \frac{1}{3} {\bm v}_j \cdot {\bm v}_\ell,
\label{eq:vec_rule}
\end{equation}
where ${\bm v}_j$ and ${\bm v}_\ell$ can be either $c$-number or operator valued vectors with three components (see also Lemma 3.3 in Ref \cite{KLT}). 
Now let us consider the correspondence between the vertex configurations and the factors appearing in the expansion of equation (\ref{eq:MVBS}). 
The factors $\sigma^a_j \, \Omega^a_{2j-1}$ and $\Omega^a_k \, \Omega^a_{k+1}$ may be expressed graphically by bonds with the color $a$ ($a=1,2,3$).  One sees from the integration rules in equation (\ref{eq:rules}) that the only diagrams which survive are those in which any vertex in the lattice ($k=1,2,...,2N-1, 2N$ in Fig. \ref{fig: VBS} (b)) is covered by zero or two bonds with the same color. With the identification $0 \leftrightarrow$ (empty bond) and $a \leftrightarrow$ ($a$-colored bond), this coincides with the condition for the vertex configurations with nonvanishing Boltzmann weights shown in Fig. \ref{fig: weights} (a). 
Then, using the rules (\ref{eq:rules}) or (\ref{eq:vec_rule}) repeatedly, one finds
\begin{equation}
M_{\rm VBS} = M ( {1}/{3}, {1}/{\sqrt 3} ).
\end{equation}
Therefore, the Gram matrix $M_{\rm VBS}$ is included in $M(x,y)$ as a limiting case.  

The above observation leads us to consider the following two-parameter deformation of $M_{\rm VBS}$:
\begin{equation}
\!\!\!\!\!\!\!\!\!\!
M_{\rm VBS} (x, y) = \int \left( \prod^{2N}_{k=1} \frac{d {\bm \Omega}_k}{4 \pi} \right)
\prod^N_{j=1} (1+{\sqrt 3} y\, {\bm \sigma}_j \cdot {\bm \Omega}_{2j-1}) 
\prod^{2N}_{k=1} (1+3 x\, {\bm \Omega}_k \cdot {\bm \Omega}_{k+1}), 
\label{eq:integralM}
\end{equation}
where the factor $\sqrt{3}$ in front of $y$ takes into account the fact that $({\bm \sigma}_j)^2 =3$ for all $ j$. This integral form of $M_{\rm VBS} (x, y)$ shows a strong similarity to the partition sum of the O($3$) loop model on a hexagonal lattice \cite{Santos, Nienhuis, Guo}. In the loop-model language, the parameters $x$ and $y$ can be thought of as the fugacities of covered bonds. Note that the fugacity of the loops is fixed as $n=3$. 
The procedure to carry out the integrals over ${\bm \Omega}_k$ in equation (\ref{eq:integralM}) is quite similar to that of a high temperature expansion for the O($n$) loop model. Again using the correspondence between the diagrams and the factors appearing in the product in equation (\ref{eq:integralM}), it is straightforward to show that 
\begin{equation}
M_{\rm VBS} (x,y) = M (x,y),
\end{equation}
i.e., the matrix product operators introduced in equation (\ref{eq:cons}) are exactly identical to the integral form equation (\ref{eq:integralM}). 

%%%%%%%%%%%%%%%%%%%%%%%%%%%%%%%%%%%%%%%%%%%%%%%%%%%%%%%%%
\section{Integrability of $M(x,y)$ associated with the unit circle}
\subsection{Properties and Proposition}

In the previous section, we have introduced two kinds of MPOs: one is $T(\lambda)$ the transfer matrix of the XXX chain, and the other is $M(x,y)$ which is identical to $M_{\rm VBS} (x,y)$ in equation (\ref{eq:integralM}). So far these two MPOs have nothing to do with each other. However, we will uncover an unexpected link between them for the case where $(x,y)$ lie on the unit circle centered at the origin. In such a case,  it is more convenient to write the parameters $(x, y)$ in terms of $\theta$ as $(x,y)=(\cos \theta, \sin \theta)$, where $\theta\in \mathbb{R}$ or can be restricted to lie in $[0,\,2\pi]$. From symbolic and numerical calculations, we find the following properties of $M(x,y)$:

\bigskip

\noindent
\textbf{Property 1.}~~ {\it For any $N$ and arbitrary $\theta \in \mathbb{R}$, $\lambda \in \mathbb{C}$, the operator $M(\cos\theta, \sin\theta)$ commutes with the transfer matrix $T(\lambda)$.} 

\medskip

\noindent
\textbf{Property 2.}~~ {\it For any $N$ and arbitrary $\theta_1, \theta_2 \in \mathbb{R}$, 
the operators $M(\cos \theta_1, \sin \theta_1)$ and $M(\cos \theta_2, \sin \theta_2)$ 
commute with each other.} 

\bigskip
\noindent
For $N \le 5$ one can, in fact, prove these properties by explicitly writing out $M (x,y)$ in terms of two trivial conserved and mutually commuting quantities, i.e., the Heisenberg Hamiltonian $H$ and the total spin operator ${\bm \sigma}_{\rm tot}$ (see Eq. (\ref{eq:tot}) for definition). In Appendix A, we show how $M(x,y)$ can be expressed as polynomials in $H$ and $({\bm \sigma}_{\rm tot})^2$. For $N>5$, $M(x,y)$ cannot be expressed only in terms of $H$ and ${\bm \sigma}_{\rm tot}$. However, we have checked numerically up to $N=13$ that Properties 1 and 2 hold within numerical accuracy for various choices of the parameters. These properties strongly suggest the existence of some integrable structures of $M(\cos \theta, \sin \theta)$, which are similar to those of the XXX chain. We will show in the following that the above properties are corollaries of the following

\bigskip

\noindent
\textbf{Proposition.}~~ {\it For any $N$ and arbitrary $\theta \in \mathbb{R}$, the operator $M(\cos\theta, \sin\theta)$ is written in terms of the transfer matrices $T$ as
\begin{equation}
M(\cos\theta, \sin\theta) = (\sin \theta)^{2N}\, T (\lambda_\theta) T (-\lambda_\theta),
\label{eq:prop}
\end{equation}
where the argument $\lambda_\theta$ is given as a function of $\theta$ by
\begin{equation}
\lambda_\theta = \frac{i}{2}\, f(\theta),
\label{eq:def_lam}
\end{equation}
with
\begin{equation}
f(\theta)=\sqrt{\frac{1+3 \cos^2 \theta}{\sin^2 \theta}}.
\end{equation}
}

\bigskip
\noindent
Note that the RHS of Eq. (\ref{eq:prop}) is even in $\lambda_\theta$, because it is invariant under sending $\lambda_\theta \to -\lambda_\theta$, which is a consequence of the commutativity of transfer matrices, i.e., Eq. (\ref{eq:TT}). Therefore, $\lambda_\theta$ obtained by replacing the RHS of Eq. (\ref{eq:def_lam}) by $-i f(\theta)/2$ gives the same $M(\cos\theta, \sin\theta)$. 
We also note that $T(-\lambda)= (-1)^N T^{t}(\lambda)$ for arbitrary $\lambda$, where ${}^{t}$ denotes matrix transpose. 
We will give a proof of Proposition in the next subsection. But before turning to the proof, let us see how Properties 1 and 2 follow as corollaries of Proposition. 
By combining Proposition with the commutativity of $T$s (Eq. (\ref{eq:TT})), it is easy to see that $M(\cos\theta, \sin\theta)$ commutes with $T(\lambda)$ for arbitrary $\lambda \in \mathbb{C}$, namely, Property 1. From Property 1, one readily sees that $M (\cos\theta, \sin\theta)$ commutes with the Hamiltonian $H$ and all other conserved charges of the XXX chain, which are derived from the logarithmic derivatives of $T(\lambda)$ (see Eq. (\ref{eq:higher_charge})).
Property 2 can also be proved by noting that $T(\lambda_{\theta_1})$, $T(-\lambda_{\theta_1})$, $T(\lambda_{\theta_2})$, and $T(-\lambda_{\theta_2})$ are commuting among themselves. 
It is interesting to ask how the operator $M(\cos\theta, \cos\theta)$ or its logarithmic derivatives with respect to $\theta$ can be expressed in terms of $H$ and higher Hamiltonians $Q_n$. It is, however, highly nontrivial and out of the scope of the present work to obtain the explicit expressions. We leave this for future work. 

\subsection{Proof of Proposition}

To prove Proposition, we first rewrite the product of two transfer matrices as a single MPO with bond dimension $D=4$. From the basic properties of tensor products, namely, 
$(A\otimes B) (C \otimes D) = (AC) \otimes (BD)$ and
${\rm Tr}[ {\sf A} ]\, {\rm Tr}[{\sf B}] = {\rm Tr} [ {\sf A} \otimes {\sf B} ]$, we have
\begin{eqnarray}
\!\!\!\!\!\!\!\!\!\!
T (\lambda) T (-\lambda) &=&   
\sum^3_{\al_1, ..., \al_N=0} \, \sum^3_{\be_1, ..., \be_N=0} 
 {\rm Tr} [ 
({\sf L}^{\al_1}(\lambda) \otimes {\sf L}^{\be_1}(-\lambda))
\cdots
({\sf L}^{\al_N}(\lambda) \otimes {\sf L}^{\be_N}(-\lambda))]\nonumber\\
&&\times 
(\sig^{\al_1} \sig^{\be_1}) \otimes 
\cdots \otimes 
(\sig^{\al_N} \sig^{\be_N}).
\label{eq:TT1}
\end{eqnarray}
Then from the properties of the identity and Pauli matrices, 
\begin{equation}
\sig^0 \sig^0 = \sig^0,~~~
\sig^0 \sig^a = \sig^a \sig^0 = \sig^a,~~~
\sig^a \sig^b = \delta^{ab} +i \epsilon^{abc} \sig^c,
\end{equation}
($a,b,c=1,2,3$), we see that Eq. (\ref{eq:TT1}) can be cast into the form
\begin{equation}
T (\lambda) T (-\lambda) 
=\sum^3_{\al_1, ..., \al_N=0}
{\rm Tr} [ {\widetilde {\sf M}}^{\al_1}(\lambda) \cdots {\widetilde {\sf M}}^{\al_N}(\lambda) ]\,
\sig^{\al_1} \otimes \cdots \otimes \sig^{\al_N},
\label{eq:TT2}
\end{equation}
where new matrices living in the $D=4$ auxiliary space are given by 
\begin{equation}
{\widetilde {\sf M}}^0 (\lambda) = \sum^3_{\al=0} {\sf L}^\al (\lambda) \otimes {\sf L}^\al (-\lambda),
\end{equation}
and, for $a=1,2,3$, 
\begin{equation}
\!\!\!\!\!\!\!\!\!\!\!\!\!\!\!
{\widetilde {\sf M}}^a (\lambda) = {\sf L}^0 (\lambda) \otimes {\sf L}^a (-\lambda) 
+{\sf L}^a (\lambda) \otimes {\sf L}^0 (-\lambda)
+i \sum^3_{b=1} \sum^3_{c=1} \epsilon^{abc} {\sf L}^b (\lambda) \otimes {\sf L}^c (-\lambda).
\end{equation}

Let us now show that the MPO with ${\widetilde {\sf M}}^\alpha (\lambda_\theta)$ is proportional to $M(\cos\theta, \sin\theta)$. To this end, we recall the fact that a matrix product operator does not have a unique representation and is, in fact, invariant under so-called gauge transformations in the auxiliary space. This comes from the cyclic property of the trace. Namely, in Eq. (\ref{eq:TT2}), the replacement  
${\widetilde {\sf M}}^{\alpha} (\lambda) \to {\sf X}^{-1} {\widetilde {\sf M}}^{\alpha}(\lambda) {\sf X}$ ($\al=0,1,2,3$) 
leaves the MPO unchanged. Therefore, if there exists an invertible matrix ${\sf X}$ such that 
${\sf X}^{-1} {\widetilde {\sf M}}^{\alpha}(\lambda_\theta) {\sf X}$ is proportional to ${\sf M}^\al (\cos\theta, \sin\theta)$ for all $\al$, then the identity (\ref{eq:prop}) holds and Proposition follows. After tedious but straightforward matrix manipulation, 
we find that the following 
\begin{equation}
{\sf X} = \frac{1}{\sqrt 2} 
\left(\begin{array}{cccc} 
0 & -1 & i & 0 \\
1 & 0 & 0 & 1 \\
-1 & 0 & 0 & 1 \\
0 & 1 & i & 0
\end{array}\right)
\left(\begin{array}{cccc} 
g(\theta) & 0 & 0 & 0 \\
0 & 1 & 0 & 0 \\
0 & 0 & 1 & 0 \\
0 & 0 & 0 & 1
\end{array}\right),
\end{equation}
with
\begin{equation}
g (\theta) = \sqrt{\frac{f(\theta) +1}{f(\theta)-1}},
\label{eq:gfunc}
\end{equation}
satisfies the demand:
\begin{equation}
(\sin\theta)^2\, {\sf X}^{-1} {\widetilde {\sf M}}^\al (\lambda_\theta) {\sf X} 
= {\sf M}^\al (\cos\theta, \sin\theta),~~~\al=0,1,2,3.
\label{eq:gauge_tr}
\end{equation}
Note that $f(\theta) \ge 1$ for all $\theta$, and thus the argument of square root in Eq. (\ref{eq:gfunc}) is nonnegative for all $\theta$. The existence of ${\sf X}$ with the property Eq. (\ref{eq:gauge_tr}) immediately implies Eq. (\ref{eq:prop}). This proves Proposition. 

\section{Summary and outlook}
We have introduced a two-parameter family of matrix product operators 
of bond dimension $D=4$, acting on the space of states of the spin-$1/2$ chain.  
We found that the operators constructed, $M(x,y)$, interpolate between 
two different categories of exactly solvable models. On the one hand, 
$M(x,y)$ is related to the Affleck-Kennedy-Lieb-Tasaki model, 
the ground states of which are called valence-bond-solid states. 
We have shown that $M(x,y)$ with $x=1/3$ and $y=1/\sqrt{3}$  
is proportional to the square root of the reduced density matrix of the 
valence-bond-solid state on a hexagonal ladder. On the other hand, 
we showed that $M(x,y)$ with $(x,y)$ satisfying $x^2+y^2=1$ is 
related to the isotropic spin-$1/2$ Heisenberg (XXX) chain, 
which is solvable by the Bethe ansatz. 
In this case, we found that the operator $M(x,y)$ is proportional to 
the product of two transfer matrices, the logarithmic derivatives of which 
are related to the Hamiltonian and conserved charges of the XXX chain. 
We have found the following properties of $M(x,y)$: 
i) for arbitrary $\theta \in \mathbb{R}$, $M(\cos\theta,\sin\theta)$ 
commutes with the Hamiltonian and all conserved charges of the XXX chain, 
ii) for arbitrary $\theta_1, \theta_2 \in \mathbb{R}$, 
$M(\cos\theta_1,\sin\theta_1)$ and $M(\cos\theta_2, \sin\theta_2)$ 
are mutually commuting. These properties were proved as corollaries of Proposition, 
which was also proved as a consequence of the Yang-Baxter equation. 

Finally, we note that Proposition implies the existence of a state whose 
reduced density matrix is proportional to the square of $M(x,y)$. 
It would be very interesting to find a quantum Hamiltonian whose 
ground state is such a state. The model is not necessarily defined on a hexagonal ladder. 
From the well-known correspondence between the XXX chain and the six-vertex model, 
the most natural candidate would be the quantum six-vertex model \cite{Ardonne} 
where the ground state is a weighted superposition of states, each of which 
is labeled by an arrow configuration on a square lattice. We are hopeful that 
the integrability structures of $M(x,y)$ revealed here will shed some light on 
the entanglement properties of conformal quantum critical systems 
in two dimensions \cite{Hsu, Stephan}. 

%%%%%%%%%%%%%%%%%%%%%%%%%%%%%%%%%%%%%%%%%%%%%%%%%%%%%%%%%%%%%%%%%
%%%%% acknowledgment   %%%%%
\ack{}
The author would like to thank Anatol N. Kirillov, Tohru Koma, Vladimir E. Korepin, and Shu Tanaka for their valuable comments and suggestions. 
The author was supported in part by JSPS Grants-in-Aid for Scientific Research nos. 23740298 and 25400407.

%%%%%%%%%%%%%%%%%%%%%%%%%%%%%%%%%%%%%%%%%%%%%%%%%%%%%%%%%%%%%%%%%

\appendix
\section{$M(x,y)$ for $N \le 5$}
\label{sec:ana}

For $N \le 5$, $M(x,y)$ can be written out explicitly as a polynomial in trivially conserved quantities: higher powers of the Hamiltonian and the total spin operator defined by 
\begin{equation}
{\bm \sigma}_{\rm tot} = \sum^N_{j=1} {\bm \sig}_j. 
\label{eq:tot}
\end{equation} 
We thus see that 
(i) $M(x,y)$ commute with $T(\lambda)$ for all $\lambda$, and 
(ii) $M(x,y)$ are mutually commuting, even if $(x,y)$ do not lie on the unit circle. 
Note that we have checked with Mathematica that this nice property does not extend to $N \ge 6$. 
In this subsection, we provide explicit expressions for $M(x,y)$ for $N \le 5$. 

\begin{itemize}
\item $N=2$
\begin{equation}
M(x,y) = 1+3x^4 + x^2 y^2  H,
\end{equation}

\item $N=3$
\begin{equation}
M(x,y) = 1+3x^6 + x^2 (1+x^2) y^2  H,
\end{equation}

\item $N=4$
\begin{eqnarray}
M(x,y) &=& 1+3x^8-6x^4 y^2 (2-y^2) 
+(3x^4y^4 -x^2 (1-x^2)y^2) H  \nonumber \\
&+& \frac{1}{2}x^4 y^4 H^2 +x^4 y^2 (1-y^2) ({\bm \sigma}_{\rm tot})^2,
\end{eqnarray}

\item $N=5$
\begin{eqnarray}
M(x,y) &=& 1+3x^{10}-\frac{15}{2}x^4 (1+x^2) y^2 +15 x^6 y^4 \nonumber \\
&+& \left[ x^2 (1-x^2) (1-x^4) y^2 +\frac{1}{2}x^4 (4-15x^2) y^4 \right] H \nonumber \\
&+& \frac{1}{2} x^4 (1-2x^2) y^4 H^2 + \frac{1}{2} x^4 y^2 [ 1+x^2-(1+2x^2)y^2 ] ({\bm \sigma}_{\rm tot})^2 \nonumber \\
&+& \frac{1}{2} x^6 y^4 H ({\bm \sigma}_{\rm tot})^2.
\end{eqnarray}

\end{itemize}

For $N=4$ and $5$, one can rewrite $({\bm \sigma}_{\rm tot})^2$ in terms of 
$H$ and the second neighbor interaction:
\begin{equation}
H' := \sum^N_{j=1} {\bm \sigma}_j \cdot {\bm \sigma}_{j+2}, 
\end{equation}
which is conserved for $N=4$, $5$ but not for any higher value of $N$ \cite{Grabowski3}. 
For $N=4$, we have 
\begin{equation}
({\bm \sigma}_{\rm tot})^2 =2 H + H'+12,
\end{equation}
and similarly for $N=5$, 
\begin{equation}
({\bm \sigma}_{\rm tot})^2 =2 H + 2 H'+15.
\end{equation}

%%%%%%%%%%%%%%%%%%%%%%%%%%%%%%%%%%%%%%%%%%%%%%%%%%%%%%%%%%%%%%%%%
\section*{References}
\providecommand{\newblock}{}


\begin{thebibliography}{10}
\expandafter\ifx\csname url\endcsname\relax
  \def\url#1{{\tt #1}}\fi
\expandafter\ifx\csname urlprefix\endcsname\relax\def\urlprefix{URL }\fi
\providecommand{\eprint}[2][]{\url{#2}}

\bibitem{Li_Haldane}
Li H and Haldane F D M 2008 {\em Phys. Rev. Lett.\/} {\bf {\textbf{101}}} 010504

\bibitem{Regnault}
Regnault N, Bernevig B A and Haldane F D M 2009 {\em Phys. Rev. Lett.\/} {\bf {\textbf{103}}} 016801
\bibitem{Zozulya}
Zozulya O S, Haque M and Regnault N 2009 {\em Phys. Rev. B\/} {\bf {\textbf{79}}} 045409
\bibitem{Lauchili}
L\"auchli A M, Bergholtz E J, Suorsa J and Haque M 2010 {\em Phys. Rev. Lett.\/} {\bf {\textbf{104}}} 156404
\bibitem{Thomale}
Thomale R, Sterdyniak A, Regnault N and Bernevig B A 2010 {\em Phys. Rev. Lett.\/} {\bf {\textbf{104}}} 180502
\bibitem{Chandran}
Chandran A, Hermanns M, Regnault N and Bernevig B A 2011 {\em Phys. Rev. B\/} {\bf {\textbf{84}}} 205136
\bibitem{Qi_Ludwig}
Qi X-L, Katsura H and Ludwig A W W 2012 {\em Phys. Rev. Lett.\/} {\bf {\textbf{108}}} 196402
\bibitem{Turner}
Turner A M, Zhang Y and Vishwanath A 2010 {\em Phys. Rev. B\/} {\bf {\textbf{82}}} 241102
\bibitem{Fidkowski}
Fidkowski L 2010 {\em Phys. Rev. Lett.\/} {\bf {\textbf{104}}} 130502
\bibitem{Prodan}
Prodan E, Hughes T L and Bernevig B A 2010 {\em Phys. Rev. Lett.\/} {\bf {\textbf{105}}} 115501
\bibitem{Calabrese}
Calabrese P and Lefevre A 2008 {\em Phys. Rev. A\/} {\bf {\textbf{78}}} 032329
\bibitem{Pollmann1}
Pollmann F and Moore J E 2010 {\em New J. Phys.\/} {\bf {\textbf{12}}} 025006
\bibitem{Pollmann2}
Pollmann F, Turner A M, Berg E and Oshikawa M 2010 {\em Phys. Rev. B\/} {\bf {\textbf{81}}} 064439
\bibitem{Poilblanc}
Poilblanc D 2010 {\em Phys. Rev. Lett.\/} {\bf {\textbf{105}}} 077202
\bibitem{Thomale2}
Thomale R, Arovas D P and Bernevig B A 2010 {\em Phys. Rev. Lett.\/} {\bf {\textbf{105}}} 116805
\bibitem{Lauchli2}
L\"auchli A M and Schliemann J 2012  {\em Phys. Rev. B\/} {\bf {\textbf{85}}} 054403
\bibitem{Yao}
Yao H and Qi X-L 2010 {\em Phys. Rev. Lett.\/} {\bf {\textbf{105}}} 080501
\bibitem{Poilblanc1}
Cirac J I, Poilblanc D, Schuch N and Verstraete F 2011 {\em Phys. Rev. B\/} {\bf {\textbf{83}}}  245134
\bibitem{Huang}
Huang C-Y and Lin F-L 2011 {\em Phys. Rev. B\/} {\bf {\textbf{84}}} 125110
\bibitem{Lou_PRB}
Lou J, Tanaka S, Katsura H and Kawashima N 2011 {\em Phys. Rev. B\/} {\bf {\textbf{84}}} 245128.
\bibitem{Tanaka1}
Tanaka S 2013 {\em Interdisciplinary Information Sciences\/} {\bf {\textbf{19}}} 101
%%%%%%%%%%%

\bibitem{Tanaka_Tamura}
Tanaka S, Tamura R and Katsura H 2012  {\em Phys. Rev. A\/} {\bf {\textbf{86}}} 032326
\bibitem{Lesanovsky1}
Lesanovsky I 2011 {\em Phys. Rev. Lett.\/} {\bf {\textbf{106}}} 025301
\bibitem{Ates}
Ji S, Ates C and Lesanovsky I 2011 {\em Phys. Rev. Lett.\/} {\bf {\textbf{107}}} 060406

\bibitem{Baxter_book}
Baxter R J 1982 {\it Exactly Solved Models in Statistical Mechanics} (Academic, London)

\bibitem{Verstraete1}
Verstraete F, Garcia-Ripoll J J and Cirac J I 2004 {\em Phys. Rev. Lett.\/} {\bf {\textbf{93}}}  207204
\bibitem{Vidal1}
Zwolak M and Vidal G 2004 {\em Phys. Rev. Lett.\/} {\bf {\textbf{93}}} 207205

\bibitem{KKKKT}
Katsura H, Kawashima N, Kirillov A N, Korepin V E and Tanaka S 2010 {\em J. Phys. A: Math. Gen.\/} {\bf {\textbf{43}}} 255303.

\bibitem{Santos}
Santos R A 2013 {\em Phys. Rev. B\/} {\bf {\textbf{87}}} 035141
\bibitem{AKLT1}
Affleck I, Kennedy T, Lieb E H and Tasaki H 1987 {\em Phys. Rev. Lett.\/} {\bf {\textbf{59}}} 799
\bibitem{AKLT2}
Affleck I, Kennedy T, Lieb E H and Tasaki H 1988 {\em Commun. Math. Phys.\/} {\bf {\textbf{115}}} 477
\bibitem{KLT}
Kennedy T, Lieb E H and Tasaki H 1988 {\em J. Stat. Phys.\/} {\bf {\textbf{53}}} 383

\bibitem{Faddeev}
Faddeev L D 1984 {\it Les Houches 1982, Recent Advances in Field Theory and Statistical Mechanics} (Elsevier Science, Amsterdam) p 561
\bibitem{Korepin_book}
Korepin V E, Bogoliubov N M and Izergin A G 1993 {\it Quantum Inverse Scattering Method and
Correlation Functions} (Cambridge University Press)
\bibitem{Nepomechie}
Nepomechie R I 1999  {\em Int. J. Mod. Phys. B\/} {\bf {\textbf{13}}} 2973 {\tt [arXiv:hep-th/9810032]}

\bibitem{Katsura_Maruyama1}
Katsura H and Maruyama I 2010 {\em J. Phys. A: Math. Gen.\/} {\bf {\textbf{43}}} 175003
\bibitem{Katsura_Maruyama2}
Maruyama I and Katsura H 2010 {\em J. Phys. Soc. Jpn.\/} {\bf {\textbf{79}}} 073002
\bibitem{Murg1}
Murg V, Korepin V E and Verstraete F 2012 {\em Phys. Rev. B\/} {\bf {\textbf{86}}} 045125

\bibitem{Znidaric1}
\v{Z}nidari\v{c} M 2010 {\em J. Phys. A: Math. Gen.\/} {\bf {\textbf{43}}} 415004
\bibitem{Prosen1}
Prosen T 2011 {\em Phys. Rev. Lett.\/} {\bf {\textbf{106}}} 217206
\bibitem{Prosen2}
Prosen T 2011 {\em Phys. Rev. Lett.\/} {\bf {\textbf{107}}} 137201 
\bibitem{Karevski} 
Karevski D, Popkov V and Sch\"utz G M 2013  {\em Phys. Rev. Lett.\/} {\bf {\textbf{110}}} 047201
\bibitem{Prosen3}
Prosen T, Ilievski E and Popkov V 2013  {\em New J. Phys.\/} {\bf {\textbf{15}}} 073051

\bibitem{Grabowski1}
Grabowski M P and Mathieu P 1994 {\em Mod. Phys. Lett.\/} {\bf {\textbf{9}}} 2197
\bibitem{Grabowski2}
Grabowski M P and Mathieu P 1995 {\em Annals Phys.\/} {\bf {\textbf{243}}} 299
\bibitem{Grabowski3}
Grabowski M P and Mathieu P 1995 {\em J. Math. Phys.\/} {\bf {\textbf{36}}} 5340
\bibitem{Tetelman}
Tetelman M G 1981 {\em Sov. Phys. JETP\/} {\bf {\textbf{55}}} 306
\bibitem{Thacker}
Thacker H B 1986 {\em Physica\/} {\bf {\textbf{18D}}} 348
\bibitem{Sklyanin}
Sklyanin E K 1992 {\it Quantum Inverse Scattering Method. Selected Topics} {\tt [arXiv:hep-th/9211111]}

\bibitem{Arovas_PRL}
Arovas D P, Auerbach A and Haldane F D M 1988 {\em Phys. Rev. Lett.\/} {\bf {\textbf{60}}} 531

\bibitem{Nienhuis}
Bl\"ote H W J and Nienhuis B 1994  {\em Phys. Rev. Lett.\/} {\bf {\textbf{72}}} 1372
\bibitem{Guo}
Guo W, Bl\"ote H W J and Wu F Y 2000 {\em Phys. Rev. Lett.\/} {\bf {\textbf{85}}} 3874

\bibitem{Ardonne}
Ardonne E, Fendley P and Fradkin E 2004 {\em Ann. Phys. (N.Y.)\/} {\bf {\textbf{310}}} 493
\bibitem{Hsu}
Hsu B, Mulligan M, Fradkin E and Kim E-A. 2009  {\em Phys. Rev. B\/} {\bf {\textbf{79}}} 115421 
\bibitem{Stephan}
St\'{e}phan J-M, Furukawa S, Misguich G and Pasquier V 2009 {\em Phys. Rev. B\/} {\bf {\textbf{80}}} 184421

\end{thebibliography}
\end{document}